\documentclass[12pt]{article}
\usepackage{graphicx}
\usepackage{lineno}
\usepackage{multirow}
\usepackage{hyperref}
\usepackage{adjustbox}

\newcommand\pubnumber{SNSN-323-63}
\newcommand\pubdate{December 4, 2022}

\def\institute{Centre for Cosmology, Particle Physics and Phenomenology (CP3),\\ Universit\`{e} Catholique Louvain, Louvain-la-Neuve, Belgium\\ and\\ Universiteit Gent, Ghent, Belgium}

\def\Title#1{\begin{center} {\Large #1 } \end{center}}
\def\Author#1{\begin{center}{ \sc #1} \end{center}}
\def\Address#1{\begin{center}{ \it #1} \end{center}}

\newcommand\pubblock{\rightline{\begin{tabular}{l} \pubnumber\\
         \pubdate  \end{tabular}}}
\newenvironment{Abstract}{\begin{quotation}  }{\end{quotation}}
\newenvironment{Presented}{\begin{quotation} \begin{center} 
             PRESENTED AT\end{center}\bigskip 
      \begin{center}\begin{large}}{\end{large}\end{center} \end{quotation}}

\usepackage{amssymb}
\usepackage{amsmath}

\newcommand{\ttbar}{\ensuremath{\text{t}\bar{\text{t}}}}
\newcommand{\PW}{\ensuremath{\text{W}}}
\newcommand{\PH}{\ensuremath{\text{H}}}
\newcommand{\PZ}{\ensuremath{\text{Z}}}
\newcommand{\Pgamma}{\ensuremath{\gamma}}

\newcommand{\ttW}{\ttbar\PW}
\newcommand{\ttWp}{\ensuremath{\ttW^{+}}}
\newcommand{\ttWm}{\ensuremath{\ttW^{-}}}
\newcommand{\ttZ}{\ttbar\PZ}
\newcommand{\ttH}{\ttbar\PH}
\newcommand{\ttg}{\ttbar\Pgamma}
\newcommand{\Pe}{\ensuremath{\text{e}}}
\newcommand{\Pmu}{\ensuremath{\mu}}
\newcommand{\intlumi}{\ensuremath{138 \, \text{fb}^{-1} }}
\newcommand{\llss}{\ensuremath{2\text{SS}\ell}}
\newcommand{\threl}{\ensuremath{3\ell}}

\newcommand{\fb}{\ensuremath{\,\text{fb}}}
\newcommand{\stat}{\ensuremath{\,\text{(stat)}}}
\newcommand{\syst}{\ensuremath{\,\text{(syst)}}}

\newcommand{\ExpXsecTTW}{\ensuremath{868\pm40\stat\pm51\syst\fb}}

\newcommand{\ExpXsecTTWp}{\ensuremath{553\pm30\stat\pm30\syst\fb}}
\newcommand{\ExpXsecTTWm}{\ensuremath{343\pm26\stat\pm25\syst\fb}}
\newcommand{\ExpXsecTTWr}{\ensuremath{1.61\pm0.15\stat\,^{+0.07}_{-0.05}\syst}}
\newcommand{\PredXsecTTWnnlo}{\ensuremath{592\,^{+155}_{-97}\,\fb}}
\newcommand{\NLONNLL}{\ensuremath{\text{NLO\,+\,NNLL}}}
\newcommand{\NLOFXFX}{\ensuremath{\text{NLO\,+\,FxFx}}}

\begin{document}

\begin{titlepage}
\pubblock
\vfill
\Title{Measurement of the cross section of top quark-antiquark pair production in association with a W boson in proton-proton collisions at $\sqrt{s} = 13$ TeV }
\vfill
\Author{ Tu Thong TRAN on behalf of the CMS collaboration}
\Address{\institute}
\vfill
\begin{Abstract}
The production of a top quark-antiquark pair in association with a W boson is measured in proton-proton collisions at a centre-of-mass energy of 13 TeV. The data recorded by the CMS experiment at the CERN LHC correspond to an integrated luminosity of 138 fb$^{-1}$. Events with two or three leptons (electrons and muons) and additional jets are selected. In events with two leptons, a multiclass neural network is used to distinguish between events from the signal and background processes. Events with three leptons are categorised based on the number of jets, number of b-tagged jets, and the lepton charges. An inclusive production cross section of $868 \pm 40 \, \text{(stat)}\pm 51 \, \text{(syst) fb}$ is measured. The cross sections of top quark-antiquark pair production with a W$^+$ and a W$^-$ boson are measured as $553 \pm 30 \, \text{(stat)} \pm 30 \, \text{(syst) fb}$ and $343 \pm 26 \, \text{(stat)} \pm 25 \, \text{(syst) fb}$, respectively, and thecorresponding ratio of the two cross sections is found to be $1.61 \pm 0.15 \, \text{(stat)}^{+0.07}_{-0.05} \, \text{(syst)}$. The results are in agreement with the standard model predictions within two standard deviations, and represent the most precise measurements of this production process to date.
\end{Abstract}
\vfill
\begin{Presented}
$15^\mathrm{th}$ International Workshop on Top Quark Physics\\
Durham, United Kingdom\\
4 -- 9 September, 2022
\end{Presented}
\vfill
\end{titlepage}
\def\thefootnote{\fnsymbol{footnote}}
\setcounter{footnote}{0}

\section{Introduction}
At the LHC, the production of a top quark-antiquark pair in association with a \PW{} boson (\ttW{}) proceeds  via quark-antiquark annihilation at leading order (LO) in quantum chromodynamics (QCD) and electroweak (EW) couplings. The absence of gluon-gluon fusion initial states at LO leads to a sizeable difference between the cross sections of $\ttW^+$ and $\ttW^-$, which results in a large charge asymmetry. Furthermore, \ttW{} is sensitive to the electroweak couplings of the top quark and it provides an experimental probe to phenomena beyond the standard model (SM). Previously, the \ttW{} production cross section has been measured at the LHC and observed higher values than predictions~\cite{CMS:TOP-17-005, ATLAS:2019fwo}. 

\begin{figure}[h!]
\centering
\includegraphics[width=0.4\textwidth]{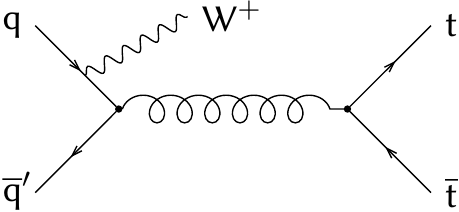}%
\hspace{0.1\textwidth}%
\includegraphics[width=0.4\textwidth]{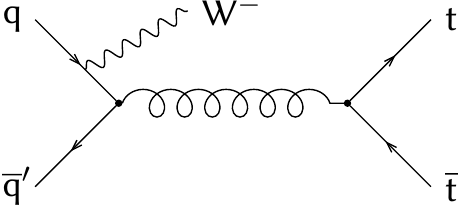} \\[0.04\textwidth]
\caption{%
    Representative Feynman diagrams of \ttW{} production at LO.
}
\label{fig:feynman}
\end{figure}

This document reports a measurement of the inclusive cross section of \ttW{} production in proton-proton collision at $\sqrt{s} = 13$ TeV~\cite{CMS:TOP-21-011}. The analysed data were recorded by the CMS detector~\cite{CMS:Detector-2008} during the CERN LHC Run II data-taking period from 2016 to 2018, corresponding to \intlumi{}. In this measurement, events with two or three charged leptons (electrons and muons), additional jets and b-tagged jets are selected. The cross sections of $\ttW^+$ and $\ttW^-$, and the ratio of the two cross sections are also measured. In this analysis, the dominant background contributions come from nonprompt leptons, the misidentification of lepton charge (misID), prompt leptons from SM processes (\ttH{}, $\ttZ/\gamma^*$, ZZ, WZ, $\ttbar\ttbar$, $\ttbar\text{VV}$, tHq, and tZq), and photon conversions.


\section{Event selection}
\begin{figure}
\centering
\includegraphics[width=0.4\textwidth]{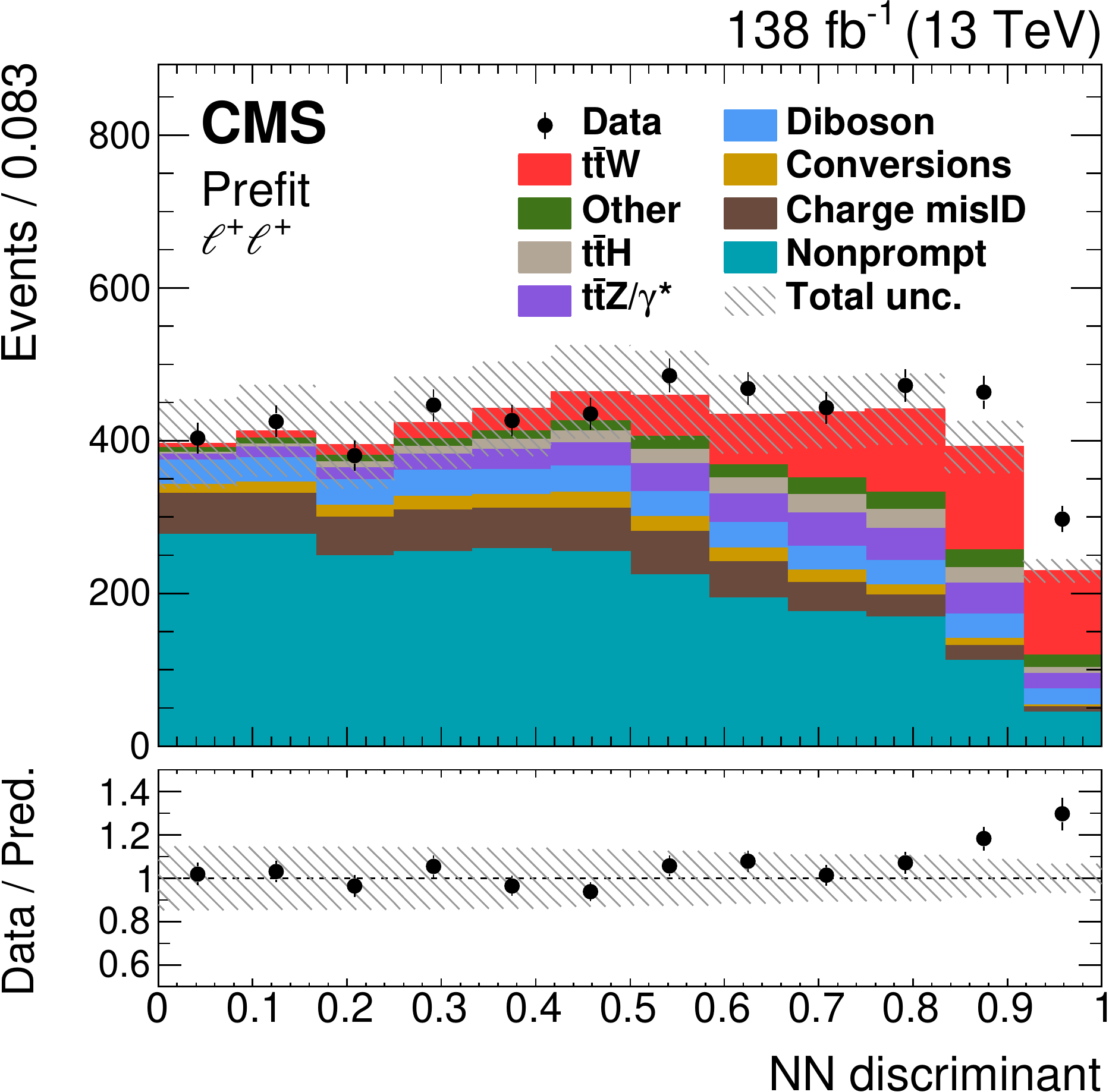}%
\hspace{0.05\textwidth}%
\includegraphics[width=0.4\textwidth]{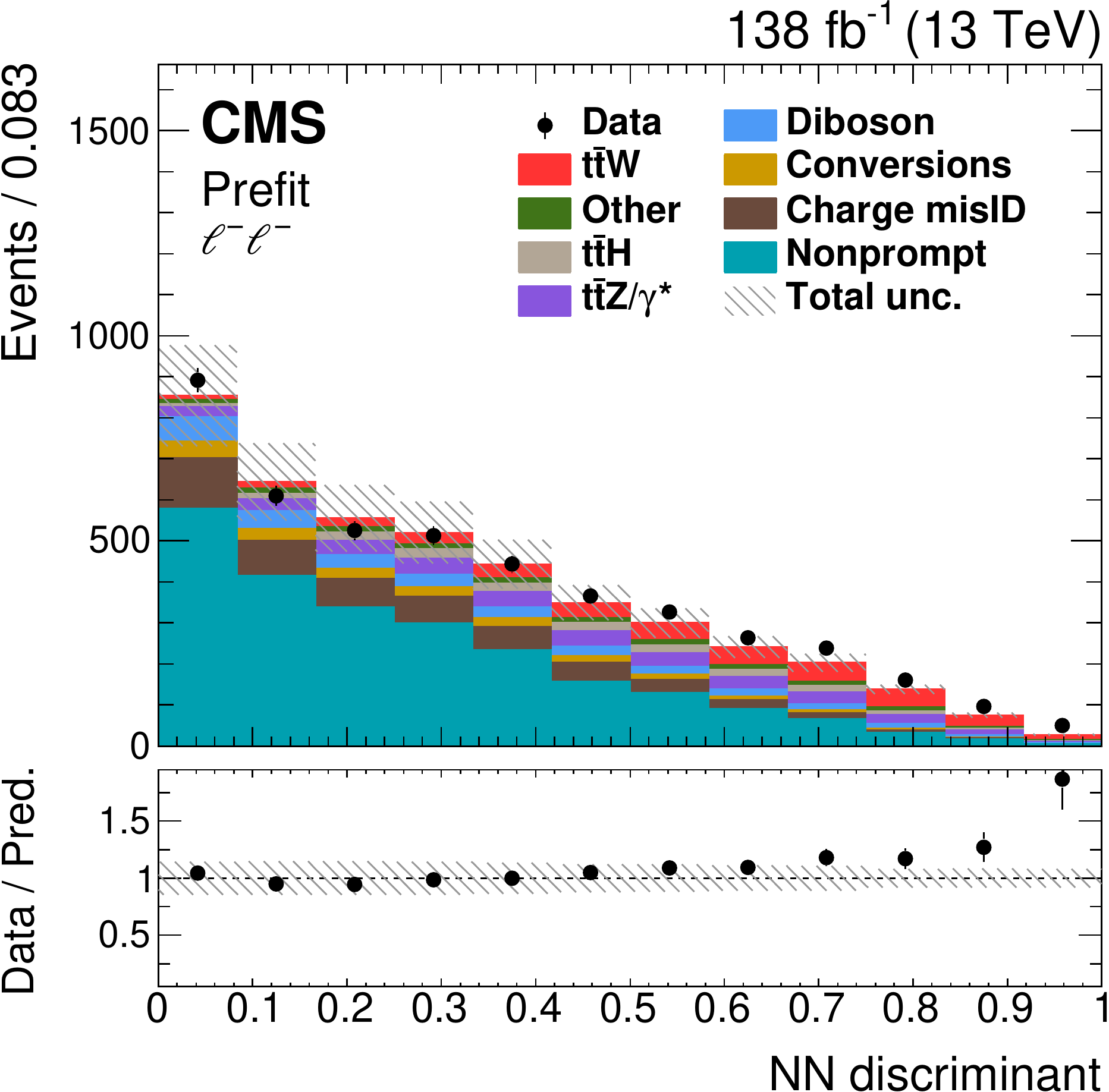} \\[1ex]
\caption{%
    Yields observed in data (points) compared with those expected from simulation (coloured histograms) as a function of NN output in signal regions with two leptons of positive (left) and negative (right) charges. The hatched bands include all systematic uncertainties in the predictions. 
    In the lower panels, the ratio of data to the sum of the predictions is shown.
}
\label{fig:fit2l}
\end{figure}

In the same-sign dilepton (\llss{}) signal region (SR), 
events are selected if they contain two tight leptons with the same charge and at least two jets, of which at least two pass the ``loose'' b tagging selection (90\% efficiency) or at least one passes the ``medium'' b tagging selection (85\% efficiency). The missing transverse momentum and invariant mass of the two leptons $m(\ell\ell)$ are required to be $>30$ GeV, and in dielectron events, $|m(\Pe\Pe) - m_\text{Z}|>15$ GeV. The two leptons must be separated by $\Delta R>0.4$. To distinguish signal from background events, a multiclass neural-network (NN) is trained using simulated \ttW{}, \ttH{}, \ttZ{}, and \ttg{} samples, and simulated \ttbar{} events as a source of nonprompt lepton. The distributions of NN output are shown in Fig.~\ref{fig:fit2l}.

In the trilepton (\threl{}) SR, events are required to have exactly three tight leptons with the sum of charges of $\pm 1$ and at least two jets, of which at least one jet must pass the ``medium'' b tagging requirements.
For any lepton pair in the event, $|m(\ell\ell)$ must be at least 12 GeV and $|m(\ell\ell) - m_\text{Z}|>10$ GeV if the two leptons are of opposite sign and same flavour. Events are classified based on the number of jets and b-tagged jets, and the invariant mass of the three leptons, $m(3\ell)$, is chosen as the discriminant. 

In addition to the two SRs above, two control regions (CR) with three or four leptons are defined. The \threl{} CR, which is enriched in ZZ, WZ and \ttZ{} processes, has a similar event selection as the \threl{}  SR, except there must be at least one same-flavour opposite-charge lepton pair with $m(\ell\ell) - m_\text{Z}| < 10 $ GeV. The four-lepton control region, which is dominated by ZZ and \ttZ{} events, is required to contain exactly four tight leptons and at least one same-flavour opposite-charge lepton pair with $|m(\ell\ell) - m_\text{Z}| < 10 $ GeV.

\section{Background estimation}
Nonprompt background contributes to both \llss{} and \threl{}. The main contribution of this background comes from \ttbar{} and it is estimated from a control sample in data using tight--to--loose method. 
Charge-misID background is relevant to \llss{} and is typically the result of the emission of a hard Bremsstrahlung photon that converses asymmetrically, hence it is negligible in dimuon final states. To estimate this background, a charge-misID rate is measured in \ttbar{} and DY simulated samples and applied to dilepton events that pass most of the event selection criteria, except the charges of the leptons are required to be opposite. 

Background contributions from prompt leptons are estimated using simulated samples, they are normalised to cross section predictions from Monte Carlo simulations. The main contributions are \ttH{}, $\ttZ/\gamma^*$, diboson (ZZ, WZ); there are also smaller contributions from VVV and rare top quark ($\ttbar\ttbar$, $\ttbar\text{VV}$, tHq, and tZq). An additional background contribution in final states with electrons arises from the conversion of a photon. This background is known as "Conversions" and is estimated using simulated \ttg{} samples.

\section{Results}
\begin{figure}[htb!]
\centering
\includegraphics[width=0.5\textwidth]{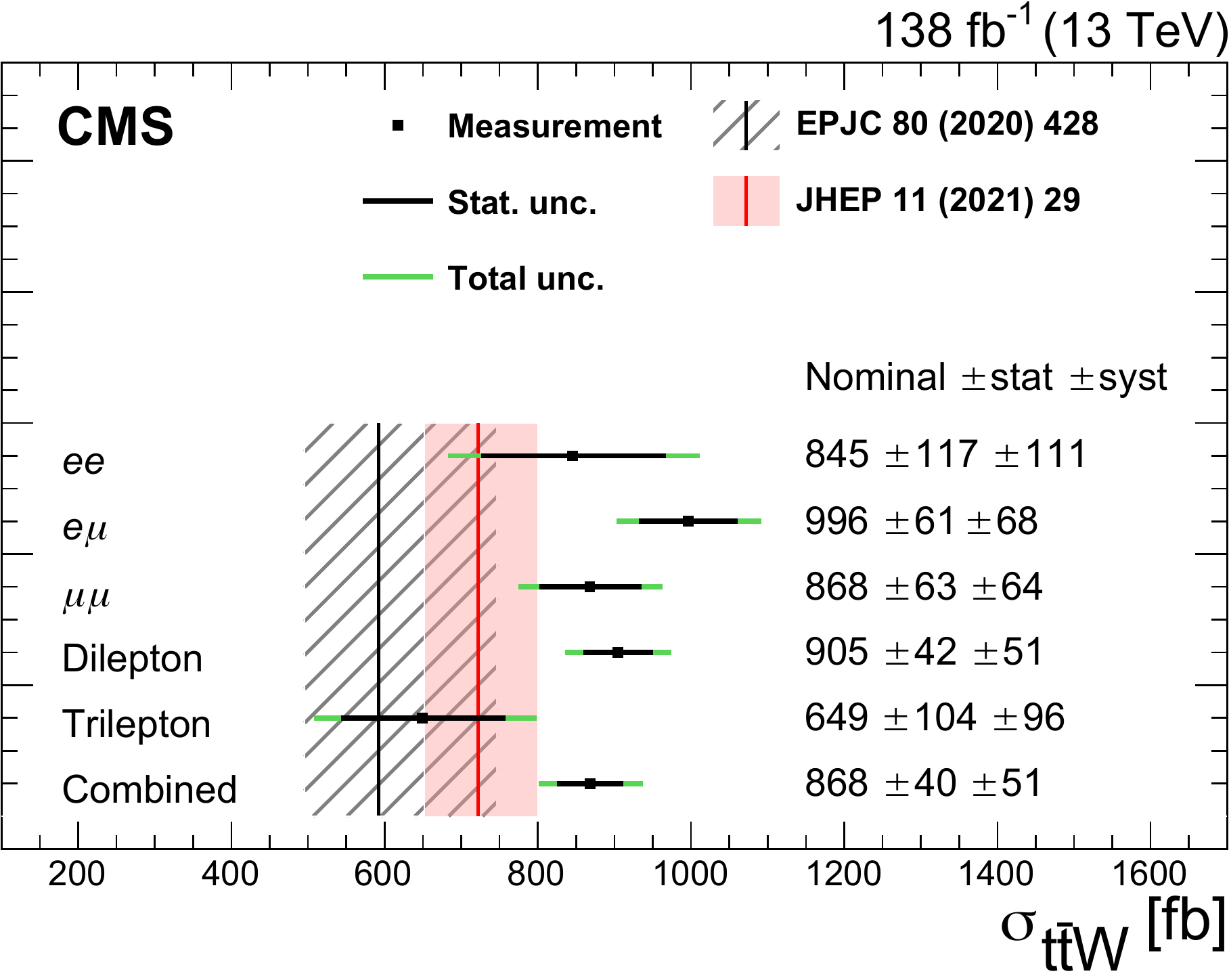}
\caption{
    A comparison the measured values of \ttW{} cross section for the individual dilepton channels ($\Pe\Pe$, $\Pe\Pmu$, $\Pmu\Pmu$, and combined) and the trilepton
    channel, as well as their combination with statistical (black bars) and total (green bars) uncertainties. 
    The predictions from two SM calculations from Refs.~\cite{Kulesza:2020nfh,
    Frederix:2021agh} are shown by the black and red vertical lines.}
\label{fig:xsecincl}
\end{figure}

By performing a binned profile likelihood fit to the NN output for the \llss{} SR, the $m(\threl)$ distributions in the \threl{} SR, and event yields from the two CRs, an inclusive cross section is obtained with a value of \ExpXsecTTW{}. The measured cross section is higher than, but consistent with the SM predictions at NLO+NNLO~\cite{Broggio:2019ewu, Kulesza:2020nfh} and NLO+FxFx~\cite{Frederix:2021agh}. This is summarised in Table~\ref{tab:results}. The dominant systematic uncertainties that are relevant to this measurement are \ttH{} normalisation, integrated luminosity, \ttW{} modelling and simulation statistical uncertainty. This result is the most precise cross section measurement of the \ttW{} production to date. A comparison of \ttW{} cross sections obtained from different final states is shown in Fig.~\ref{fig:xsecincl}.  
\begin{table}[h!]
\caption{Summary of measured and predicted values of the \ttW{}, $\ttW^+$, and $\ttW^-$ cross sections, and the ratio $\sigma_{\ttW^+}/\sigma_{\ttW^-}$.}
\begin{adjustbox}{width=0.85\columnwidth,center}
\begin{tabular}{ccccc}
    \multirow{2}{*}{Observable} & \multirow{2}{*}{Measurement} & \multicolumn{2}{c}{SM prediction} \\[-4pt]
    & & \NLONNLL~\cite{Broggio:2019ewu, Kulesza:2020nfh} & \NLOFXFX~\cite{Frederix:2021agh} \\ \hline
    $\sigma_{\ttW}$ & \ExpXsecTTW & \PredXsecTTWnnlo & $722\,^{+71}_{-78}\,\fb$ \\
    $\sigma_{\ttW^+}$ & \ExpXsecTTWp & $384\,^{+53}_{-33}\,\fb$ & $475\,^{+46}_{-52}\,\fb$ \\
    $\sigma_{\ttW^-}$ & \ExpXsecTTWm & $198\,^{+26}_{-17}\,\fb$ & $247\,^{+24}_{-27}\,\fb$ \\
    $\sigma_{\ttW^+}/\sigma_{\ttW^-}$ & \ExpXsecTTWr & $1.94\,^{+0.37}_{-0.24}\,$ & $1.92\,^{+0.27}_{-0.29}\,$ \\
\end{tabular}
\end{adjustbox}
\label{tab:results}
\end{table}

The cross sections of $\ttW^+$ and $\ttW^-$ are obtained from a simultaneous fit to data, where the events are categorised based on the sum of charges of the leptons. The measured cross sections are \ExpXsecTTWp{} and \ExpXsecTTWm{} for $\ttW^+$ and $\ttW^-$, respectively. Additionally, the measurement of the ratio between the cross sections of $\ttW^+$ and $\ttW^-$ is also reported to be $\sigma_{\ttW^+}/\sigma_{\ttW^-} = \ExpXsecTTWr$. The two cross sections are larger than, but consistent with the SM predictions while the measured ratio is smaller than the SM predictions.

\begin{figure}[htbp!]
\centering
\includegraphics[height=0.3\textwidth]{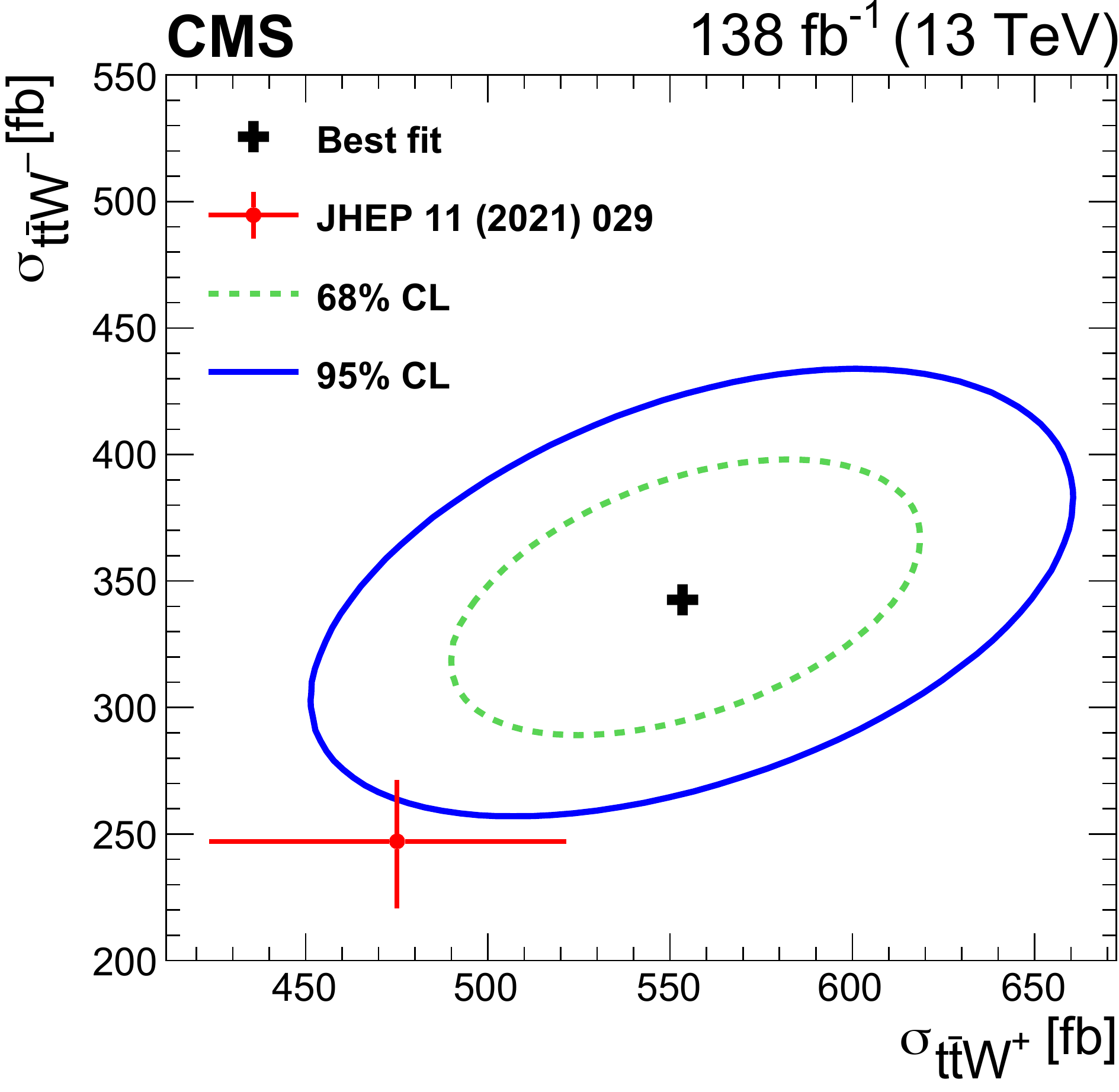}
\includegraphics[height=0.3\textwidth]{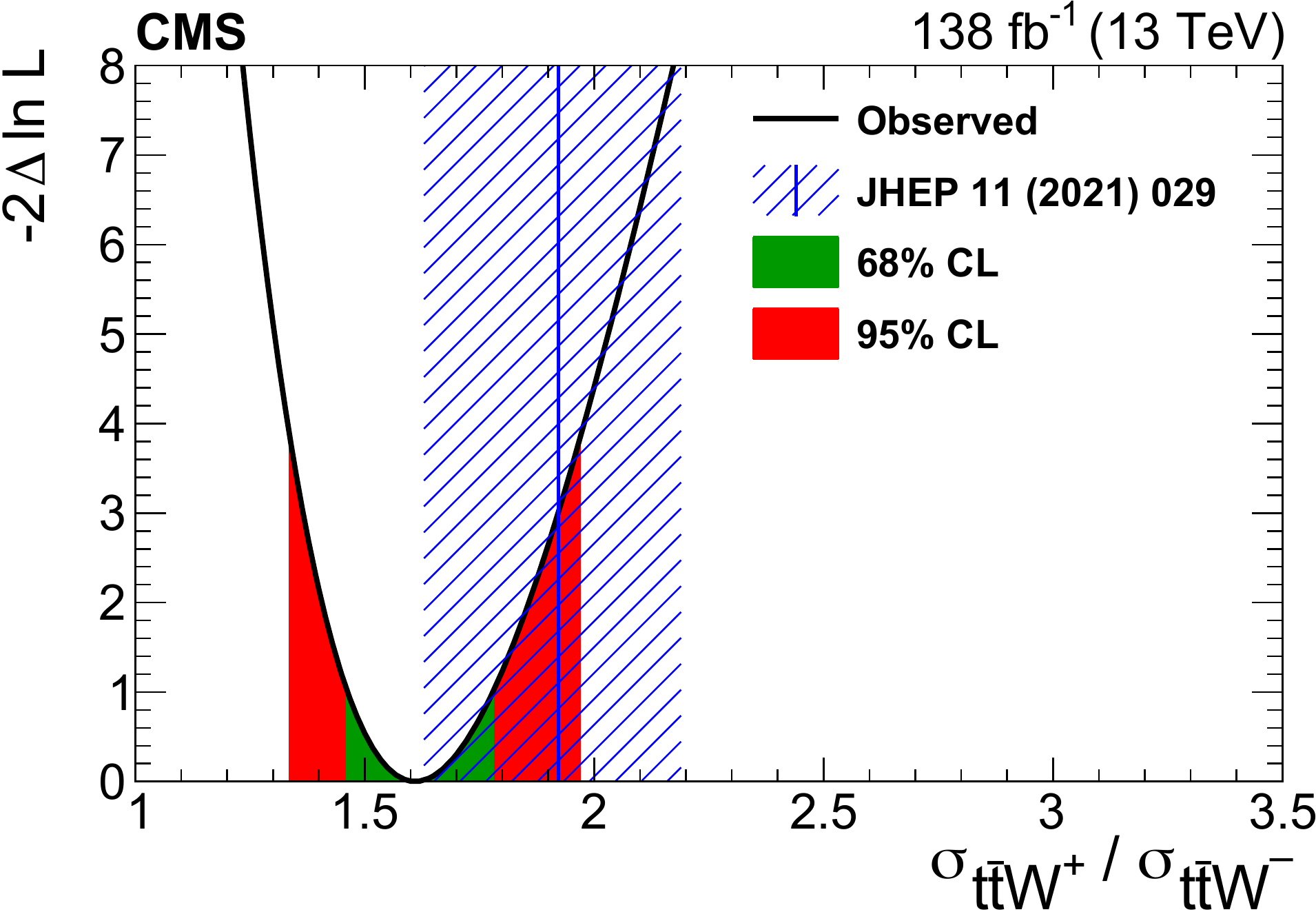}
\caption{The measurement of the \ttWp vs. \ttWm cross sections (left) and the scan of the negative log-likelihood used in the measurement of the cross section ratio $\sigma_{\ttW^+}/\sigma_{\ttW^-}$ (right) with their associated 68 and 95\% CL intervals.
The SM predictions of the cross sections provided by the authors of Ref.~\cite{Frederix:2021agh}, and the ratio is found from the cross section values.
}
\label{fig:ratio}
\end{figure}

\section*{Acknowledgement}
This work has received funding from FWO and F.R.S--FNRS under the ``Excellence of Science (EOS) programme'' no. 30820817.

\end{document}